\documentstyle[aps,version2]{revtex}

\begin{document}
\newtheorem{theorem}{Theorem}
\newtheorem{lemma}{Lemma}
\def\qed{\hfill \vrule height 7pt width 7pt depth 0pt \smallskip}
\hfill{ITD 92/92--10}\par
\draft
\begin{title}
Introduction to Random Matrices
\end{title}
\author{Craig A.~Tracy}
\begin{instit}
Department of Mathematics and Institute of Theoretical Dynamics,\\
University of California,
Davis, CA 95616, USA
\end{instit}
\author{Harold Widom}
\begin{instit}
Department of Mathematics,\\
University of California,
Santa Cruz, CA 95064, USA
\end{instit}
\begin{abstract}
These notes provide an introduction to the theory of random matrices.
The central quantity studied is $\tau(a)={\rm det}\left(1-K\right)$
where $K$ is the integral operator with kernel
\[  {1\over \pi} {\sin\pi(x-y)\over x-y}\, \chi_I(y)\, . \]
Here $I=\bigcup_j\left(a_{2j-1},a_{2j}\right)$ and $\chi_I(y)$ is the
characteristic function of the set $I$.  In the Gaussian Unitary
Ensemble (GUE) the probability that no eigenvalues lie in $I$
is equal to $\tau(a)$.  Also $\tau(a)$ is a tau-function
and we present a new simplified derivation of the system of
nonlinear completely integrable equations (the $a_j$'s are the
independent variables) that were first derived by Jimbo, Miwa,
M{\^o}ri, and Sato in 1980.  In the case of a single interval
these equations are reducible to a Painlev{\'e} V equation.
For large $s$ we give an asymptotic formula for $E_2(n;s)$,
which is the probability in the GUE that exactly $n$ eigenvalues
lie in an interval of length $s$.
\end{abstract}
\section{INTRODUCTION}
\label{sec:intro}
These notes provide  an introduction to that aspect of the theory of random
matrices dealing with   the distribution of
eigenvalues.  To first orient the reader,
 we present in Sec.~\ref{sec:numer}
some numerical experiments that illustrate some of the basic aspects of
the subject.  In Sec.~\ref{sec:inv} we introduce
 the invariant measures for the three ``circular ensembles''  involving
unitary matrices.  We also define the level spacing distributions
and express these distributions in terms of a particular Fredholm
determinant.
In Sec.~\ref{sec:ope} we explain  how these measures are modified
for the orthogonal polynomial ensembles.
  In Sec.~\ref{sec:univ} we discuss
the universality of these level spacing distribution functions in
a particular scaling limit.
The discussion up to this point (with the possible
exception of Sec.~\ref{sec:univ}) follows the well-known path
pioneered by
Hua, Wigner, Dyson, Mehta and others who first developed this theory
(see, e.g., the reprint volume of Porter~\cite{porter}
and Hua~\cite{hua}).
This, and much more, is discussed in Mehta's book~\cite{mehta_book}---{\it
the classic reference\/} in the subject.
\par
An important development in random matrices was the discovery  by
Jimbo, Miwa, M{\^o}ri, and Sato~\cite{jmms} (hereafter
referred to as JMMS)   that the basic Fredholm determinant
mentioned above is a $\tau$-function in the sense of the Kyoto School.
Though it has been some twelve years since~\cite{jmms} was published,
these  results are not widely appreciated by
the practitioners of random matrices. This is due no doubt to the
complexity of their paper.  The methods of JMMS are methods of discovery;  but
now that we know the result,  simpler proofs can be constructed.  In
Sec.~\ref{sec:jmms_eqs} we give such a proof of the JMMS
equations.  Our  proof is  a simplification
and generalization  of Mehta's~\cite{mehta92b} simplified
proof of the single interval case.
Also our methods build on the earlier work
of Its, Izergin, Korepin, and Slavnov~\cite{its90}
and Dyson~\cite{dyson92}.
We include in  this section  a discussion of the  connection
between the JMMS equations and the integrable
Hamiltonian systems  that appear in the geometry of quadrics and spectral
theory  as developed by Moser~\cite{moser}.
This section concludes  with a discussion of
the case of a  single interval (viz.,
probability that exactly $n$ eigenvalues
lie in a given interval).  In this
case  the JMMS equations can be reduced to a
single ordinary differential equation---the
 Painlev{\'e} V equation.
\par
Finally, in Sec.~\ref{sec:asym} we discuss the asymptotics
in the case of a large single interval  of
the various level spacing distribution
 functions~\cite{btw,widom92,mehta_mahoux}.
  In this analysis both the
Painlev{\'e} representation and new results in Toeplitz/Wiener-Hopf
theory are needed to produce these asymptotics.
We also give an approach based on the asymptotics of the
eigenvalues of the basic linear integral
 operator~\cite{fuchs64,mehta_book,slepian}.    These results
are then compared with the continuum model calculations of
Dyson~\cite{dyson92}.

\section{NUMERICAL EXPERIMENTS}
\label{sec:numer}
The Gaussian orthogonal ensemble (GOE)
 consists of $N\times N$ real symmetric matrices
whose elements (subject to
the symmetric constraint)  are independent and identically distributed
Gaussian random variables
of mean zero and variance one.  One can use a Gaussian random number generator
to produce a ``typical'' such matrix.  Given this matrix we can
diagonalize it  to produce our ``random eigenvalues.''
Using the software MATHEMATICA,
25 such $100\times 100$ GOE matrices were  generated
 and  Fig.~1 is    a histogram
of the density of eigenvalues
  where the $x$-axis has been normalized so that
all eigenvalues lie in $[-1,1]$.
Also shown is the
{\it Wigner semicircle law\/}
\begin{equation}
  \rho_{W}(x)={2\over \pi}\sqrt{1-x^2}.
\end{equation}
\par
Given any such distribution  (or density) function, one can ask
to what extent is it ``universal.''
In Fig.~2 we plot the same density histogram except we change
the distribution of matrix elements to the uniform distribution on
$[-1,1]$.  One sees that the same semicircle law is a good approximation
to the density of eigenvalues.  See~\cite{mehta_book}
for further discussion of the Wigner semicircle law.
\par
A fundamental quantity of the theory is the (conditional)  probability
that given  an eigenvalue at $a$,  the next
eigenvalue lies between $b$ and $b+db$: $p(0;a,b)\, db$.
In measuring this quantity it is
usually  assumed that
the system is well approximated by a translationally
invariant system of constant
eigenvalue density.  This density
is conveniently normalized to one.    In the translationally invariant
system  $p(0;a,b)$ will depend only
upon the difference $s:=b-a$.  When there is no
chance for  confusion, we denote this probability
density simply by $p(s)$.
Now  as  Figs.~1 and 2 clearly show,  the
eigenvalue density in the above examples is not constant.  However, since we
are mainly interested in the case of large matrices
(and ultimately $N\rightarrow\infty$),   we take for our
data the eigenvalues lying in an interval in which the density does not
change significantly.  For this data  we compute a histogram
of {\it spacings\/}  of eigenvalues.
That is to say, we order the eigenvalues $E_i$ and
compute the level spacings $S_i:=E_{i+1}-E_i$.
 The scale of both the $x$-axis
and $y$-axis are fixed once we require that the integrated density
is one and the mean spacing is one. Fig.~3 shows the resulting histogram
for 20,  $100\times 100$ GOE matrices where the eigenvalues were taken
from the middle half of the entire eigenvalue spectrum.  The important
aspect of this data is it shows {\it  level repulsion of eigenvalues\/}
as indicated by the vanishing of $p(s)$ for small $s$.
Also plotted is the {\it Wigner surmise\/}
\begin{equation}
p_{W}(s)= {\pi\over 2}s \exp\left( -{\pi\over 4} s^2\right) \label{wigner}
\end{equation}
which for these purposes numerically well approximates the exact result
(to be discussed below)  in the
range $0\leq s \leq 3$.  In Fig.~4 we show the same histogram  except
now the data are  from 50,   $100\times 100$ real symmetric
matrices whose elements
are iid random variables  with uniform distribution on $[-1,1]$.
\par
In computing these  histograms,  the spacings were
computed for each  realization of a random matrix, and then
the level spacings of several experiments were lumped together to form
the data.  If one first forms the data by mixing together the
eigenvalues from the random matrices, and then computes the level spacing
histogram the results are completely different.  This is
illustrated in  Fig.~5  where the resulting histogram is
well approximated by the Poisson density $\exp( -s) $ (see Appendix 2
in~\cite{mehta_book} for further discussion of the superposition
of levels).
There are other  numerical experiments that can be done, and the
reader is invited to discover various aspects of random matrix theory
by devising ones own experiments.

\section{INVARIANT MEASURES AND LEVEL SPACING DISTRIBUTIONS}
\label{sec:inv}
\subsection{Preliminary Remarks}
\label{subsec:inv_remarks}
In classical statistical mechanics, the microcanonical ensemble is defined
by the measure that assigns equal a priori probability to all states
of the  given system (which in turn
is defined by specifying a Hamiltonian on phase space)
 of fixed energy $E$  and volume $V$.
 The motivation for this measure
is that after specifying the energy of the system, every point in
phase space lying on the energy surface should be equally likely
since we have ``no further macroscopic information.''  In the applications of
random matrix theory to quantum systems,
 the Hamiltonian is modeled by a matrix $H$.
However, in this case we give up knowledge of the system, i.e.\
$H$ itself is unknown.
Depending upon the symmetries of the system, the sets of  possible $H$'s
are taken to be
 real symmetric matrices, Hermitian matrices, or self-dual Hermitian
matrices (``quaternion real'')
  \cite{dyson62a,dyson62b,mehta_book}. The question then is what measure
do we choose for each of these sets of  $H$'s?
\par
What we would intuitively  like to do is to make each
$H$ equally likely to reflect our ``total ignorance'' of the system.
Because  these spaces of $H$'s  are noncompact,
it is of course impossible to have a probability measure
that assigns equal a priori probability.
It is useful to recall the situation on the real line $\mbox{\bf R}$.
If we confine ourselves to a finite interval $[a,b]$, then the
unique translationally invariant
 probability measure is the normalized Lebesgue
measure.  Another characterization of this probability
measure
is that it is the unique density that maximizes the information entropy
\begin{equation}
S[p]=-\int_a^b p(x) \log p(x) \, dx. \label{entropy}
\end{equation}
On $\mbox{\bf R}$ the maximum entropy density subject to the constraints
$E(1)=1$ and $E(x^2)=\sigma^2$ is the Gaussian density of variance
$\sigma^2$.
The Gaussian ensembles of random matrix theory can also be characterized
as those measures that have maximum information entropy
subject to the constraint of a fixed value of $E(H^*H)$~\cite{balian,stone}.
The well-known explicit formulas are given below in Sec.~\ref{sec:ope}.
\par
Another approach,  first taken by Dyson~\cite{dyson62a}
and the one we follow here,   is to consider
unitary matrices rather than hermitian matrices.  The
advantage here is that the space of unitary matrices is compact
and the eigenvalue density is constant (translationally invariant
distributions).

\subsection{Haar Measure for $U(N)$}
\label{subsec:haar}
We denote by $G=U(N)$ the set of $N\times N$ unitary matrices and recall
that ${\rm dim}_{\mbox{\bf R}} U(N)=N^2$.  One can think of $U(N)$
as an $N^2$-dimensional submanifold of $\mbox{\bf R}^{2N^2}$ under
the identification of a complex $N\times N$ matrix with a point
in $\mbox{\bf R}^{2N^2}$.
The group $G$ acts on itself by either left or right translations, i.e.\
fix $g_0\in G$ then
\[
L_{g_0}:g\rightarrow g_{0}g \ \ \ {\rm and} \ \ \ R_{g_0}:g\rightarrow g g_0
\, .\label{translations}
\]
The normalized Haar measure $\mu_2$  is the unique probability measure
on $G$  that
is both left- and right-invariant:
\begin{equation}
\mu_2(g E)=\mu_2(E g) =\mu_2(E)\label{haar_inv}
\end{equation}
for all $g\in G$ and every measurable set $E$ (the reason
for the subscript 2 will become clear below).
Since for compact groups a measure that is left-invariant is also
right-invariant, we need only construct a left-invariant measure
to obtain the Haar measure.
The invariance (\ref{haar_inv}) reflects the precise meaning of
``total ignorance'' and is the analogue  of   translational invariance
picking  out the Lebesgue measure.
\par
To construct the Haar measure we construct the matrix of left-invariant
1-forms
\begin{equation}
\Omega_g=g^{-1} dg\label{one_forms},\ \ \ g\in G,
\end{equation}
where $\Omega_g$ is anti-Hermitian since $g$ is unitary. Choosing
$N^2$ linearly independent 1-forms $\omega_{ij}$ from $\Omega_g$,  we
form the associated volume form
obtained by taking the wedge product
of these $\omega_{ij}$'s.  This volume form on $G$ is left-invariant,
and hence up to normalization, it is the desired  probability
measure.
\par
Another way to contruct the Haar measure is to  introduce
the standard Riemannian metric on the space of $N\times N$ complex
matrices $Z=(z_{ij})$:
\[
(ds)^2={\rm tr}\left( dZ dZ^*\right)=\sum_{j,k=1}^N\left\vert dz_{ij}
\right\vert^2.
\label{metric1}
\]
We now restrict this metric to the submanifold of unitary matrices.  A
simple computation shows the restricted metric is
\begin{equation}
(ds)^2={\rm tr}\left(\Omega_g \Omega_g^*\right).\label{metric2}
\end{equation}
Since $\Omega_g$ is left-invariant, so is the metric $(ds)^2$.  If
we use the standard Riemannian formulas to construct the volume element,
we will arrive at the invariant volume form.
\par
We are interested in the induced
probability density on the  eigenvalues.  This calculation
is  classical and
can be found in~\cite{hua,weyl}.  The
derivation  is particularly clear if we
make use of the Riemannian metric (\ref{metric2}).  To this end
we write
\begin{equation}
g=X\, D \, X^{-1}\, , \ \ \ g\in G,\label{diagonal_form}
\end{equation}
where $X$ is a unitary matrix and $D$ is a diagonal matrix whose diagonal
elements we write as $\exp(i\varphi_k)$.
Up to an ordering of the angles $\varphi_k$ the matrix $D$ is unique.
We can assume that the eigenvalues are distinct since  the degenerate
case has  measure zero.  Thus  $X$ is determined up to a
diagonal unitary matrix.  If we denote by $T(N)$ the subgroup
of diagonal unitary matrices, then to each $g\in G$ there corresponds
a unique pair $(X,D)$, $X\in G/T(N)$ and $D\in T(N)$.  The Haar
measure induces a measure on $G/T(N)$ (via the natural projection
map). Since
\[
X^* dg X=\Omega_X D - D \Omega_X + dD\, ,
\]
we have
\begin{eqnarray*}
(ds)^2&=&{\rm tr}\left(\Omega_g\Omega_g^*\right)={\rm tr}\left(dg\,dg^*\right)
={\rm tr}\left(X^* dg X X^* dg^* X\right) \\
&=&{\rm tr}\left([\Omega_X,D]\,[\Omega_X,D]^*\right)+
{\rm tr}\left(dD\, dD^*
\right) \\
& =& \sum_{k,\ell=1}^N \left\vert \delta X_{k\ell}\left(\exp(i\varphi_k)-
\exp(i\varphi_\ell)\right)\right\vert^2 + \sum_{k=1}^N(d\varphi_k)^2
\end{eqnarray*}
where $\Omega_X=(\delta X_{k\ell})$.  Note that the diagonal elements
$\delta X_{kk}$ do not appear in the metric.  Using the Riemannian
volume formula we obtain the volume form
\begin{equation}
\omega_g=\omega_X \prod_{j<k}\left\vert \exp(i\varphi_j)-\exp(i\varphi_k)
\right\vert^2
d\varphi_1\cdots d\varphi_{N}
\end{equation}
where $\omega_X={\rm const} \prod_{j>k}\delta X_{jk}$.
\par
We now integrate over the entire group $G$ subject to the condition
that the elements have their angles between $\varphi_k$ and
$\varphi_k+d\varphi_k$ to obtain
\begin{theorem}
The volume of that part of the unitary group $U(N)$  whose elements have
their angles between $\varphi_k$ and $\varphi_k+d\varphi_k$ is given
by
\begin{equation}
P_{N 2}(\varphi_1,\ldots,\varphi_N) \,
d\varphi_1\cdots d\varphi_{N}=C_{N 2}\prod_{j<k}\left\vert
\exp(i\varphi_j)-\exp(i\varphi_k)\right\vert^2 \,  d\varphi_1\cdots d\varphi_N
\end{equation}
where $C_{N 2}$ is a normalization constant.
\end{theorem}
\subsection{Orthogonal and Symplectic Ensembles}
\label{subsec:beta14}
Dyson~\cite{dyson62b},
 in a careful analysis of the implications of time-reversal
invariance for physical systems, showed that (a)  systems having
time-reversal invariance and rotational symmetry or having
time-reversal invariance and integral spin are characterized by
symmetric unitary matrices; and (b) those systems having time-reversal
invariance and half-integral spin
but  no rotational symmetry are characterized by self-dual unitary
matrices (see also Chp.~9 in~\cite{mehta_book}).
For systems without  time-reversal invariance there  is  no restriction on
the unitary matrices.  These three sets of unitary matrices along
with their respective invariant measures, which we denote
by $E_\beta(N)$, $\beta=1,4,2$, respectively,   constitute the {\it circular
ensembles\/}.  We denote the invariant measures by $\mu_\beta$, e.g.\
$\mu_2$ is the normalized Haar measure discussed in Sec.~\ref{subsec:haar}.
\par
A symmetric unitary matrix $S$ can be written as
\[
S=V^T V,  \ \ \ V\in U(N).
\]
Such a decomposition is not unique since
\[
V\rightarrow R\,  V, \ \ \ R\in O(N),
\]
leaves $S$ unchanged.
Thus the space of symmetric unitary matrices can be identified with
the coset space
\[
 U(N)/O(N) \, .
\]
The group $G=U(N)$ acts on the coset space $U(N)/O(N)$,  and we want
the invariant measure $\mu_1$.  If $\pi$ denotes the natural
projection map $G\rightarrow G/H $, then the measure $\mu_1$ is
just the induced measure:
\[
\mu_1(B) = \mu_2(\pi^{-1}(B))\, ,
\]
and hence $E_1(N)$ can be identified with the pair
$(U(N)/O(N),\mu_1)$.
\par
The space of self-dual unitary
matrices   can be identified (for even $N$) with the coset
space
\[
U(N)/Sp(N/2)
\]
where $Sp(N)$ is the symplectic group.
Similarly, the circular ensemble $E_4(N)$ can be identified with
$(U(N)/Sp(N/2),\mu_4)$ where $\mu_4$ is the induced measure.
\par
As in Sec.~\ref{subsec:haar} we want  the
the probability density $P_{N \beta}$  on the eigenvalue angles that
results from the measure $\mu_\beta$ for  $\beta=1$ and
$\beta=4$. This calculation is somewhat more involved and we refer
the reader to the original literature~\cite{dyson62a,hua}  or to Chp.~9 in
\cite{mehta_book}.  The basic result is
\begin{theorem}
\label{theorem:prob_density}
In the ensemble $E_\beta(N)$ ($\beta=1,2,4$)
 the probability of finding the eigenvalues
$\exp(i\varphi_j)$ of $S$ with an angle in each of the intervals
$\left(\theta_j,\theta_j + d \theta_j\right)$ ($j=1,\ldots,N$) is given
by
\begin{equation}
P_{N\beta}(\theta_1,\ldots,\theta_N) \, d\theta_1\cdots d\theta_N
=C_{N\beta} \prod_{1\leq \ell<j\leq N} \left\vert \exp(i\theta_\ell)
-\exp(i\theta_j)\right\vert^\beta \, d\theta_1\cdots d\theta_N
\end{equation}
where $C_{N\beta}$ is a normalization constant.
\end{theorem}
The normalization constant follows from
\begin{theorem}
\label{theorem:partition_fn}
Define for $N\in\mbox{\bf  N}$ and $\beta\in\mbox{\bf  C}$
\begin{equation}
\Psi_N(\beta)=(2\pi)^{-N} \int_0^{2\pi}\cdots \int_0^{2\pi}
\prod_{j<k}\left\vert\exp(i\theta_j)-\exp(i\theta_k)\right\vert^\beta\,
d\theta_1\cdots d\theta_N
\label{selberg}
\end{equation}
then
\begin{equation}
\Psi_N(\beta)={\Gamma(1+\beta N/2)\over \left(\Gamma(1+\beta/2)\right)^N}\, .
\end{equation}
\end{theorem}
The integral (\ref{selberg})  has an interesting history,  and it is
now understood to be a special case of Selberg's integral (see, e.g., the
discussion in~\cite{mehta_book}).
\par
\subsection{Physical Interpretation of the Probability Density
 $P_{N \beta}$}
\label{subsec:coulomb}
The 2D Coulomb potential for  $N$ unit like charges on a circle of radius
 one is
\begin{equation}
W_N(\theta_1,\ldots,\theta_N)
=-\sum_{1\leq j<k\leq N} \log\left\vert \exp(i\theta_j)-\exp(i\theta_k)
\right\vert\, .
\end{equation}
Thus the (positional) equilibrium Gibbs measure at
inverse temperature $0<\beta <\infty$ for
 this system of  $N$ charges with energy $W_N$ is
\begin{equation}
{\exp\left(-\beta W_N(\theta_1,\ldots,\theta_N)\right)\over \Psi_N(\beta)}\, .
\end{equation}
For the special cases of  $\beta=1,2,4$
this Gibbs measure is the probability density of
 Theorem~\ref{theorem:prob_density}.  Thus in this mathematically
equivalent description, the term ``level repulsion'' takes on the
physical meaning of repulsion of charges. This Coulomb gas description
is due to  Dyson~\cite{dyson62a},  and it suggests various
physically natural  approximations that
would not otherwise  be so clear.
\subsection{Level Spacing Distribution Functions}
\label{subsec: level_spacing}
For large matrices there is too much information in the probability
densities $P_{N\beta}(\theta_1,\ldots,\theta_N)$ to be useful in
comparison with data.  Thus we want to integrate out some of this
information.  Since the probability density $P_{N\beta}(\theta_1,\ldots,
\theta_N)$ is a completely symmetric function of its arguments, it
is natural to introduce the {\it $n$-point correlation functions\/}
\begin{equation}
R_{n \beta}(\theta_1,\ldots,\theta_n)={N!\over (N-n)! }
\int_0^{2\pi}\cdots\int_0^{2\pi} P_{N \beta} (\theta_1,\ldots,\theta_N)\,
d\theta_{n+1}\cdots d\theta_N\, . \label{n_point}
\end{equation}
The case $\beta=2$ is significantly simpler to handle and we limit
our discussion here to this case.
\bigbreak\bigbreak
\begin{lemma}
\begin{equation}
P_{N 2}(\theta_1,\ldots,\theta_N)={1\over N!}\det\left( K_N(\theta_j,
\theta_k)\Bigr\vert_{j,k=1}^N\right)
\end{equation}
where
\begin{equation}
K_N(\theta_j,\theta_k)= {1\over 2\pi} {\sin\left(N
(\theta_j-\theta_k)/2\right)\over \sin\left((\theta_j-\theta_k)/2\right)}\> .
\label{S_function}
\end{equation}
\end{lemma}
Proof:
\par
Recalling the Vandermonde determinant we can write
\begin{equation}
\prod_{j<k}\left\vert \exp(i\theta_j)-\exp(i\theta_k)\right\vert^2
=\det(M^T)\det(\overline M)
\end{equation}
where $M_{jk}=\exp\left(i(j-1)\theta_k\right)$.  A simple calculation
shows
\begin{equation}
\left(M^T \overline M\right)_{jk} =
2\pi D_{jj} K_N(\theta_j,\theta_k) \bar D_{kk}
\end{equation}
where $D$ is the diagonal matrix with entries
 $\exp\left(i(N-1)\theta_j/2\right)$. Except for the normalization
constant the lemma now follows.
Getting the correct normalization constant requires a little more work
(see, e.g.,  Chp.~5 in~\cite{mehta_book}). \qed
\par
{}From this lemma and  the combinatoric Theorem~5.2.1 of
 \cite{mehta_book} follows
\begin{theorem}
\label{theorem:n_point}
Let $R_{n2}(\theta_1,\ldots,\theta_n)$ be the $n$-point function
defined by (\ref{n_point}) for the circular ensemble $E_2(N)$;  then
\begin{equation}
R_{n2}(\theta_1,\ldots, \theta_n)=\det\left(K_N(\theta_j,\theta_k)
\Bigr\vert_{j,k=1}^n\right)
\end{equation}
where $K_N(\theta_j,\theta_k)$ is given by (\ref{S_function}).
\end{theorem}
\par
We now discuss the behavior for large $N$.
The 1-point correlation function
\begin{equation}
R_{1,2}(\theta_1)= {N\over 2\pi}
\end{equation}
is just the density, $\rho$, of eigenvalues with
mean spacing $D=1/\rho$.  As the size of the
matrices goes to infinity so does the density.  Thus if we are
to construct a meaningful limit as $N\rightarrow\infty$ we must
scale the angles $\theta_j$.  This motivates
 the definition of the  {\it scaling limit\/}
\begin{equation}
\rho\rightarrow\infty, \ \  \theta_j\rightarrow 0, \ \ \mbox{\rm
such that}\ \  x_j:=\rho\theta_j\in\mbox{{\bf R}} \ \ \mbox{\rm is \ fixed}.
\label{scaling_limit}
\end{equation}
We will abbreviate this scaling limit by simply writing $N\rightarrow\infty$.
In this limit
\begin{equation}
R_{n2}(x_1,\ldots,x_n)\, dx_1\cdots dx_n :=
\lim_{N\rightarrow\infty}
R_n(\theta_1,\ldots,\theta_n) \, d\theta_1\cdots d\theta_n
\end{equation}
where we used the slightly confusing notation of denoting the scaling
limit of the $n$-point functions by the same symbol.  From
Theorem~\ref{theorem:n_point} follows
\begin{theorem}
In the scaling limit (\ref{scaling_limit}) the $n$-point functions
become
\begin{equation}
R_{n2}(x_1,\ldots,x_n)=\det\left(K(x_j,x_k)\Bigr\vert_{j,k=1}^n\right)
\end{equation}
where the kernel $K(x,y)$ is given by
\begin{equation}
K(x,y)= {1\over \pi}\,  {\sin\pi(x-y)\over x-y} \> .
\label{sine_kernel}
\end{equation}
\end{theorem}
The three  sets
 of correlation functions
\begin{equation}
 {\cal E}_\beta:=\left\lbrace R_{n \beta}(x_1,\ldots,x_n);
x_j\in\mbox{\bf R}
\right\rbrace_{n=1}^\infty,\ \ \beta=1,2,4,
\label{ensembles}
\end{equation}
define  three
different  statistics
(called the {\it orthogonal ensemble, unitary ensemble,
symplectic ensemble\/}, respectively)
  of an infinite  sequence of eigenvalues (or as we sometimes
say, {\it   levels\/}) on the real line.
\par
We now have the necessary machinary to discuss the level spacing
correlation functions in the ensemble ${\cal E}_2$.  We denote
by ${\cal I}$ the union of $m$ disjoint sub-intervals of the unit
circle:
\begin{equation}
{\cal I}={\cal I}_1\cup\cdots\cup {\cal I}_m.
\end{equation}
We begin with
the probability of finding exactly $n_1$ eigenvalues in interval
${\cal I}_1$,\ldots, $n_m$ eigenvalues in interval ${\cal I}_m$
in the ensemble $E_2(N)$.  We denote
this probability by $E_{2 N}(n_1,\ldots,n_m;{\cal I})$ and we will
let $N\rightarrow\infty$ at the end.
\par
If $\chi_A$ denotes the characteristic function of the set $A$
and $n:=n_1+\cdots +n_m$, then
the probability we want is
\begin{eqnarray}
E_{2 N}(n_1,\ldots,n_m;{\cal I})&= &{N\choose n_1\cdots n_m \,  N-n}
\int_0^{2\pi}d\theta_1\cdots\int_0^{2\pi}
d\theta_N \,  P_{N2}(\theta_1,\ldots,\theta_N) \nonumber\\
&&\times\prod_{j_1=1}^{n_1}\chi_{{\cal I}_1}(\theta_{j_1})
\prod_{j_2=n_1+1}^{n_1+n_2}
\chi_{{\cal I}_2}(\theta_{j_2})\cdots \prod_{j_m=n_1+\cdots+1}^{n_1+\cdots+n_m}
\chi_{{\cal I}_m}(\theta_{j_m})\nonumber\\
&&\times
\prod_{j=n+1}^N\left(1-\chi_{\cal I}(\theta_j)\right)\> .
\label{prob1}
\end{eqnarray}
We define the quantities
\begin{eqnarray}
r_{n_1\cdots n_m}&=&\int_0^{2\pi}d\theta_1\cdots\int_0^{2\pi}d\theta_n\,
R_n(\theta_1,\ldots,\theta_n)\nonumber\\
&&\times
\prod_{j_1=1}^{n_1}\chi_{{\cal I}_1}(\theta_{j_1})\prod_{j_2=n_1+1}^{n_1+n_2}
\chi_{{\cal I}_2}(\theta_{j_2})\cdots \prod_{j_m=n_1+\cdots+1}^{n_1+\cdots+n_m}
\chi_{{\cal I}_m}(\theta_{j_m})\> .
\end{eqnarray}
The idea is  to expand the last product term in (\ref{prob1}) involving
the characteristic function $\chi_{{\cal I}}$ and to  regroup the
terms according to the number of times a factor of $\chi_{{\cal I}}$ appears.
Doing this one can then integrate out those $\theta_k$ variables
that do not appear
as arguments   of any of  the characteristic functions and express
the result in terms of the quantities $r_{n_1\cdots n_m}$.  To recognize
the resulting terms we define the Fredholm determinant
\[
D_N({\cal I};\lambda)=\det\left(1-
\sum_{j=1}^m \lambda_j K_N(\theta,\theta^\prime)
\chi_{{\cal I}_j}(\theta^\prime)\right)
\]
where ``$\sum_{j=1}^m \lambda_j K_N(\theta,\theta^\prime)
\chi_{{\cal I}_j}(\theta^\prime)$'' means the operator with that kernel
and $\lambda$ is the $m$-tuple $(\lambda_1,\ldots,\lambda_m)$.
A slight rewriting of the Fredholm expansion gives
\[
D_N({\cal I};\lambda)=1+\sum_{j=1}^\infty (-1)^j
\sum_{{{j_k\geq 0}
 \atop j_1+\cdots+j_m=j}}{\lambda^{j_1}\cdots\lambda^{j_m}\over
j_1!\cdots j_m!}\, r_{n_1\cdots n_m} \> .
\]
The expansion above is then recognized to be proportional to
\[
{\partial^n D_N({\cal I};\lambda)\over \partial\lambda_1^{n_1}\cdots
\partial\lambda_m^{n_m}}\Bigr\vert_{\lambda_1=\cdots=\lambda_m=1}\> .
\]
The scaling limit $N\rightarrow\infty$ can be taken with the result:
\begin{theorem}
\label{theorem:fredholm_rep}
Given  $m$ disjoint open  intervals
$I_k=\left(a_{2k-1},a_{2k}\right)\subset\mbox{ {\bf R}}$, let
\begin{equation}
I:=I_1\cup\cdots\cup I_m.
\end{equation}
The probability
$E_2(n_1,\ldots,n_m;I)$ in the ensemble ${\cal E}_2$
 that exactly $n_k$ levels occur in interval $I_k$
($k=1,\ldots,m$) is given
by
\begin{equation}
E_2(n_1,\ldots,n_m,I)=
{(-1)^n\over n_1!\cdots n_m!}{\partial^n D(I;\lambda)\over
\partial\lambda^{n_1}\cdots\partial\lambda^{n_m}}
\Bigr\vert_{\lambda_1=\cdots=\lambda_m=1}
\label{level_spacing_distr}
\end{equation}
where
\begin{equation}
D(I;\lambda)=\det\left(1-\sum_{j=1}^m \lambda_j K(x,y)\chi_{I_j}(y)\right)
,\label{fredholm_det}
\end{equation}
$K(x,y)$ is given by (\ref{sine_kernel}), and
$n:=n_1+\cdots +n_m$.
\end{theorem}
\par
In the case of a single interval $I=(a,b)$,  we write
 the probability in ensemble ${\cal E}_\beta$  of exactly $n$
eigenvalues in $I$   as $E_\beta(n;s)$ where
$s:=b-a$.
Mehta~\cite{mehta_book,mehta92a} has shown that if we
define
\begin{equation}
D_\pm(s;\lambda)=\det\left(1-\lambda K_{\pm}\right)
\label{det_pm}
\end{equation}
where $K_\pm$ are  the operators  with kernels $K(x,y)\pm K(-x,y)$,
and
\begin{equation}
E_\pm(n;s)={(-1)^{n}\over n!}\, {\partial^n D_\pm(s;\lambda)\over
\partial\lambda^n}\Bigr\vert_{\lambda=1},
\end{equation}
then
$E_1(0;s)=E_+(0;s)$,
\begin{equation}
E_+(n;s)=E_1(2n;s)+E_1(2n-1;s),\ \ \ n>0\; , \label{EP}
\end{equation}
\begin{equation}
E_-(n;s)=E_1(2n;s)+E_1(2n+1;s),\ \ \ n\geq 0\; ,\label{EM}
\end{equation}
and
\begin{equation}
 E_4(n;s)=\frac{1}{ 2}\bigl ( E_+(n;2s)+E_-(n;2s)\bigr),\ \ \
n\geq 0\; .\label{E4}
\end{equation}
Using the Fredholm expansion,  small $s$ expansions can be found
for $E_\beta(n;s)$.  We quote~\cite{mehta_book,mehta92a}
 here only the results for $n=0$:
\begin{eqnarray}
E_1(0;s)&=& 1-s +{\pi^2 s^3\over 36} - {\pi^4 s^5\over 1200} +\mbox{\rm O}(s^6)
\> ,\nonumber\\
E_2(0;s)&=&1-s+{\pi^2 s^4\over 36}-{\pi^4 s^6\over 675} +\mbox{\rm O}(s^8)
\> , \nonumber\\
E_4(0;s)&=&1-s+{8\pi^4 s^6\over 2025}+\mbox{\rm O}(s^8)\> .
\label{E_small}
\end{eqnarray}
The conditional probability
in the ensemble ${\cal E}_\beta$
 of an eigenvalue between $b$ and $b+db$ given
an eigenvalue at $a$, is given~\cite{mehta_book} by
$p_\beta(0;s)\, ds$ where
\begin{equation}
p_\beta(0;s)={d^2 E_\beta(0;s)\over ds^2}\> .
\label{p_E}
\end{equation}
Using this formula and the expansions (\ref{E_small}) we see
that $p_\beta(0;s)=\mbox{\rm O}(s^\beta)$, making connection with
the numerical results discussed in Sec.~\ref{sec:numer}. Note that the
Wigner surmise (\ref{wigner})  gives for
small $s$  the correct power of $s$ for $p_1(0;s)$, but
the slope is incorrect.

\section{ORTHOGONAL POLYNOMIAL ENSEMBLES}
\label{sec:ope}
\par
Orthogonal polynomial ensembles have been studied since the 1960's,
e.g.~\cite{fox_kahn}, but recently interest has revived because
of their application to the matrix models of 2D quantum
gravity~\cite{brezin_kazakov,douglas_shenker,gross_migdal,gross}.
Here we give the main results that generalize the previous sections.
\par
The orthogonal polynomial ensemble associated to $V$ assigns a
probability measure on the space of $N\times N$ hermitian matrices
proportional to
\begin{equation}
\exp\left(-{\rm Tr}\left(V(M)\right)\right)\, dM
\end{equation}
where $V(x)$ is a real-valued function such that
\[ w_V(x)=\exp\left(-V(x)\right) \]
defines a weight function in the sense of orthogonal polynomial
theory.  The quantity $dM$ denotes the product of Lebesgue measures
over the independent elements of the hermitian matrix $M$.  Since
${\rm Tr}\left(V(M)\right)$ depends only upon the eigenvalues of $M$,
we may diagonalize $M$
\[ M=X D X^* \, , \]
and as before integrate over the ``$X$'' part of the measure
to obtain a probability measure on the eigenvalues.  Doing this
gives the density
\begin{equation}
\label{ope_eigenvalues}
P_N(x_1,\ldots,x_N)=\prod_{1\leq k< \ell \leq N} (x_k-x_\ell)^2
\exp\left(-\sum_{j=1}^N V(x_j)\right) \, .
\end{equation}
\par
If we introduce the orthogonal polynomials
\begin{equation}
\int_{\bf R} p_m(x) p_n(x) w_V(x)\, dx = \delta_{mn},\ \ \ m,n=0,1,\ldots
\end{equation}
and associated functions
\[\varphi_m(x)=\exp\left(-V(x)/2\right) p_m(x)\, , \]
then the probability density (\ref{ope_eigenvalues}) becomes
\begin{eqnarray}
P_N(x_1,\ldots,x_N)&=& {1\over N!}\, \left({\rm det}\left (
\varphi_{j-1}(x_k)\right)\bigr\vert_{j,k=1}^N\right)^2 \nonumber \\
&=& {1\over N!}\, {\rm det}\left( K_N(x_j,x_k)\right)
\end{eqnarray}
where
\begin{eqnarray}
K_N(x,y)&=& \sum_{j=0}^{N-1} \varphi_j(x)\varphi_j(y) \nonumber \\
&=& {k_{N-1}\over k_N}\> {\varphi_N(x) \varphi_{N-1}(y) -
\varphi_{N-1}(x)\varphi_N(y)\over x-y} \ \ \ {\rm for}\ \ x\neq y
\nonumber \\
&=& {k_{N-1}\over k_N}\>\left( \varphi^\prime_N(x)\varphi_N(x)-
\varphi_{N-1}^\prime(x) \varphi_N(x)\right) \ \ \ {\rm for} \ \ x=y.
\label{ope_kernel}
\end{eqnarray}
The last two equalities follow from  the Christoffel-Darboux formula and the
$k_n$ are defined by
\[ p_n(x) = k_n x^n +\cdots ,\ \ k_n>0. \]
\par
Using the orthonormality
of the $\varphi_j(x)$'s one shows exactly as in~\cite{mehta_book}
that
\begin{eqnarray}
R_n(x_1,\ldots,x_n):&=&{N!\over (N-n)!}\, \int_{\bf R} \cdots
\int_{\bf R} P_N(x_1,\ldots,x_N)\, dx_{n+1}\cdots dx_N \nonumber \\
   &=& {\rm det}\left(K_N(x_j,x_k)\bigr\vert_{j,k=1}^n\right)\, .
\end{eqnarray}
In particular, the density of eigenvalues is given by
\begin{equation}
\rho_N(x) = K_N(x,x)\, .
\label{ope_density}
\end{equation}
\par
Arguing as before, we find the probability that an interval $I$
contains no eigenvalues is given by the determinant
\begin{equation}
{\rm det}\left(1-K_N\right)
\end{equation}
where $K_N$ denotes the operator with kernel
\[ K_N(x,y) \chi_I(y) \]
and  $K_N(x,y)$ given by (\ref{ope_kernel}).
Analogous formulas hold for the probability of exactly $n$ eigenvalues
in an interval.  We remark that the size of the matrix $N$ has been
kept fixed throughout this discusion.
\section{UNIVERSALITY}
\label{sec:univ}
We now consider the limit as the size of the matrices tends to
infinity in the orthogonal polynomial ensembles of Sec.~\ref{sec:ope}.
Recall that we defined the scaling limit by introducing new variables
such that in these scaled variables the mean spacing of eigenvalues
was unity.  In Sec.~\ref{sec:inv} the system for finite $N$ was
translationally invariant and had constant mean density $N/2\pi$.
The orthogonal polynomial ensembles are not translationally invariant
(recall (\ref{ope_density}) so we now take a {\it fixed\/} point $x_0$
in the support of $\rho_N(x)$
and examine the local statistics of the eigenvalues in some small
neighborhood of this point $x_0$.
Precisely, the scaling limit in the orthogonal polynomial ensemble
that we consider is
\begin{equation}
N\rightarrow\infty,\ \ x_j\rightarrow x_0 \ \ \ {\rm such\ that} \ \
\xi_j:=\rho_N(x_0) (x_j-x_0)\ \ {\rm is\ fixed.}
\label{ope_scaling}
\end{equation}
\par
The problem is to compute $K_N(x,y)\, dy$ in this scaling limit.
{}From (\ref{ope_kernel}) one sees this is a question of asymptotics
of the associated orthogonal polynomials.  For weight functions
$w_V(x)$ corresponding to classical orthogonal polynomials such
asymptotic formulas are well known and using these it has been
shown~\cite{fox_kahn,nagao_wadati} that
\begin{equation}
K_N(x,y)\, dy \rightarrow {1\over\pi}\, {\sin\pi (\xi-\xi^\prime)\over
\xi-\xi^\prime}\, d\xi^\prime\, .
\label{kernel_limit}
\end{equation}
Note this result is independent of $x_0$ and any of the parameters
that might appear in the weight function $w_V(x)$.  Moore~\cite{moore}
has given heuristic semiclassical arguments that show that we can
expect (\ref{kernel_limit}) in great generality.
\par
There is a growing literature on asymptotics of orthogonal polynomials
when the weight function $w_V(x)$ has polynomial $V$, see~\cite{lubinsky87}
and references therein.  For example, for the case of $V(x)=x^4$
Nevai~\cite{nevai84} has given rigorous asymptotic formulas for
the associated polynomials.  It is straightforward to use Nevai's
formulas to verify that (\ref{kernel_limit}) holds in this
non-classical case.
\par
There are places where (\ref{kernel_limit}) will fail and these
will correspond to choosing the $x_0$ at the ``edge of the
spectrum'' \cite{bowick_brezin,moore} (in these cases $x_0$ varies
with $N$).  This will correspond to the double scaling limit
discovered in the matrix models of 2D quantum gravity
\cite{brezin_kazakov,douglas_shenker,gross_migdal,gross}.
 Different multicritical points
will have different limiting $K_N(x,y)\, dy$~\cite{bowick_brezin,moore}.
\bigbreak\bigbreak
\section{JIMBO-MIWA-M{\^O}RI-SATO EQUATIONS}\nobreak
\label{sec:jmms_eqs}
\subsection{Definitions and Lemmas}
\label{subsec:jmms_lemmas}
\nopagebreak[3]
\par
In this section we denote by $a$ the $2m$-tuple $(a_1,\ldots,a_{2m})$
where the $a_j$ are the endpoints of the intervals given in
Theorem \ref{theorem:fredholm_rep} and by $d_a$ exterior differentiation
with respect to the $a_j$ ($j=1,\ldots,2m$).  We make the
specialization
\[ \lambda_j=\lambda\ \ \ {\rm for}\ \ j \]
which is the case consider in~\cite{jmms}.  It is not difficult
to extend the considerations of this section to the general case.
\par
We denote by $K$ the operator that has kernel
\begin{equation}
\lambda K(x,y) \chi_I(y)
\end{equation}
where $K(x,y)$ is given by (\ref{sine_kernel}).  It is convenient
to write
\begin{equation}
\lambda K(x,y) = {A(x) A^\prime(y)-A^\prime(x) A(y)\over x-y}
\label{A_kernel}
\end{equation}
where
\[  A(x) = {\sqrt\lambda\over\pi}\sin\pi x\, . \]
The operator $K$ acts on $L^2({\bf R})$, but can be restricted
to act on a dense subset of smooth functions.  From calculus we
get the formula
\begin{equation}
{\partial\over\partial a_j} K = (-1)^j K(x,a_j)\delta(y-a_j)
\label{K_derivative}
\end{equation}
where $\delta(x)$ is the Dirac delta function.  Note that in the
right hand side of the above equation we are using the shorthand
notation that ``$A(x,y)$'' means the operator with that kernel.
We will continue to use this notation throughout this section.
\par
We introduce the functions
\begin{equation}
Q(x;a) = (1-K)^{-1} A(x) = \int_{\bf R} \rho(x,y) A(y)\, dy
\label{Q}
\end{equation}
and
\begin{equation}
P(x;a)=(1-K)^{-1} A^\prime(x) = \int_{\bf R} \rho(x,y) A^\prime(y)\, dy
\label{P}
\end{equation}
where $\rho(x,y)$ denotes the distributional kernel of $(1-K)^{-1}$.
We will sometimes abbreviate these to $Q(x)$ and $P(x)$, respectively.
It is also convenient to introduce the resolvent kernel
\[ R=(1-K)^{-1} K\, .\]
In terms of kernels, these are related by
\[ \rho(x,y)= \delta(x-y)+R(x,y)\, . \]
\par
We define the fundamental 1-form
\begin{equation}
\omega(a):=d_a\log D(I;\lambda)\, .
\label{omega}
\end{equation}
Since the integral operator $K$ is trace-class and depends smoothly
on the parameters $a$, we have the well known result
\begin{equation}
\omega(a)=-{\rm Tr}\left((1-K)^{-1} d_aK\right)\, .
\label{omega2}
\end{equation}
Using (\ref{K_derivative}) this last trace can be expressed in terms
of the resolvent kernel:
\begin{equation}
\omega(a)=-\sum_{j=1}^{2m}(-1)^j R(a_j,a_j)\, da_j
\label{omega3}
\end{equation}
which shows the importance of the quantities $R(a_j,a_j)$.
A short calculation establishes
\begin{equation}
{\partial\over\partial a_j} (1-K)^{-1}=(-1)^j R(x,a_j) \rho(a_j,y)\, .
\label{resolvent_derivative}
\end{equation}
\par
We will need two commutators which we state as lemmas.
\bigbreak\bigbreak
\begin{lemma}
\label{lemma:d_commutator}
If $D={d\over dx}$  denotes the differentiation operator, then
\[ \left[D,(1-K)^{-1}\right]=-\sum_{j=1}^{2m} (-1)^j  R(x,a_j) \rho(a_j,y)\,
.\]
\end{lemma}
Proof:
\par
Since
\[ \left[D,(1-K)^{-1}\right] = (1-K)^{-1}\left[ D,K\right] (1-K)^{-1}\, ,
\]
we begin by computing $[D,K]$.  An integration by parts shows that
\[ \left[D,K\right]=-\sum_j (-1)^j K(x,a_j)\delta(y-a_j) \]
where we used the property
\[ {\partial K(x,y)\over\partial x}+{\partial K(x,y)\over\partial y}=0\]
satisfied by our $K(x,y)$ and the well known formula for the
derivative of $\chi_I(x)$.  The lemma now follows from the fact
that
\[ R(x,y)=\int_{\bf R} \rho(x,z) K(z,y)\, dz = \int_{\bf R} K(x,z)
\rho(z,y)\, dz\, . \] \qed
\par
\bigbreak\bigbreak
\begin{lemma}
\label{lemma:x_commutator}
If $M_x$ denotes the multiplication by $x$ operator, then
\[ \left[M_x,(1-K)^{-1}\right]= Q(x)\>\left(1-K^t\right)^{-1} A^\prime
\chi_I(y)-P(x)\> \left(1-K^t\right)^{-1} A \chi_I(y) \]
where $K^t$ denotes the transpose of $K$, and also
\[ \left[M_x,\left(1-K\right)^{-1}\right]=(x-y)R(x,y)\, .\]
\end{lemma}
Proof:
\par
We have
\[ \left[ M_x,K\right]=\left( A(x) A^\prime(y)-A^\prime(x) A(y)\right)
\chi_I(y) \left(1-K\right)^{-1} \, .\]
{}From this last equation the first part of the lemma follows using
the definitions of $Q$ and $P$.  The alternative expression for the
commutator follows directly from the definition of $\rho(x,y)$ and
its relationship to $R(x,y)$.\qed
\par
\bigbreak\bigbreak
This lemma leads to a representation for the kernel $R(x,y)$~\cite{its90}:
\begin{lemma}
\label{lemma:R_representation}
If $R(x,y)$ is the resolvent kernel of (\ref{A_kernel}) and $Q$
and $P$ are defined by (\ref{Q}) and (\ref{P}), respectively, then
for $x,y\in I$ we have
\[ R(x,y) = {Q(x;a)P(y;a)-P(x;a) Q(y;a)\over x-y}\, , \ \ x\neq y\, ,\]
and
\[ R(x,x) = {dQ\over dx}(x;a)\,  P(x;a)- {dP\over dx}(x;a)\,  Q(x;a)\, . \]
\end{lemma}
Proof:
\par
Since   $K(x,y)=K(y,x)$ we have, on $I$,
\[ (1-K^t)^{-1} A \chi_I = (1-K)^{-1} A \chi_I = (1-K)^{-1} A \]
(the last since the kernel of $K$ vanishes for $y\not\in I$).
Thus $(1-K^t)^{-1} A \chi_I = Q$ on $I$, and similarly
$(1-K^t)^{-1}A^\prime \chi_I = P$ on $I$.  The first part
of the lemma then follows from  Lemma \ref{lemma:x_commutator}.
  The expression for the diagonal follows from
Taylor's theorem.\qed
\par
We remark that Lemma \ref{lemma:R_representation} used only the
property that the kernel $K(x,y)$ can be written as (\ref{A_kernel})
and not the specific form of $A(x)$.  Such kernels are called
``completely integrable integral operators'' by Its et.~al.~\cite{its90}
and Lemma \ref{lemma:R_representation} is central to their work.
\subsection{Derivation of the JMMS Equations}
\label{subsec:jmms_derivation}
We set
\begin{equation}
q_j=q_j(a)=\lim_{x\rightarrow a_j\atop x\in I}Q(x;a) \ \ {\rm and}
\ \ p_j=p_j(a) = \lim_{x\rightarrow a_j\atop x\in I}P(x;a),\ \ j=1,\ldots,2m.
\label{q_and_p}
\end{equation}
Specializing Lemma \ref{lemma:R_representation} we obtain
immediately
\begin{equation}
R(a_j,a_k) = {q_j p_k - p_j q_k \over a_j-a_k}\, ,\ \ j\neq k\, .
\label{jmms1}
\end{equation}
Referring to (\ref{resolvent_derivative}) we easily deduce that
\begin{equation}
{\partial q_j \over \partial a_k} = (-1)^k R(a_j,a_k) q_k\, ,\ \ j\neq k\, ,
\label{jmms2}
\end{equation}
and
\begin{equation}
{\partial p_j \over \partial a_k} = (-1)^k R(a_j,a_k) p_k\, ,\ \ j\neq k\,.
\label{jmms3}
\end{equation}
\par
Now
\begin{eqnarray}
{dQ\over dx}&=& D\left(1-K\right)^{-1} A(x) \nonumber \\
&=& \left(1-K\right)^{-1} D A(x) + \left[ D,\left(1-K\right)^{-1}\right] A(x)
\nonumber \\
&=& \left(1-K\right)^{-1} A^\prime(x) - \sum_k(-1)^k R(x,a_k) q_k\, .
\nonumber
\end{eqnarray}
Thus
\begin{equation}
{dQ\over dx}(a_j;a)=p_j-\sum_k(-1)^k R(a_j,a_k) q_k\, .
\label{dQ/dx}
\end{equation}
Similarly,
\begin{eqnarray}
{dP\over dx}&=& \left(1-K\right)^{-1} A^{\prime\prime}(x) +
\left[ D,\left(1-K\right)^{-1}\right] A^\prime(x) \nonumber \\
&=& -\pi^2 \left(1-K\right)^{-1} A(x) - \sum_k(-1)^k R(x,a_k) p_k\, .
\nonumber
\end{eqnarray}
\bigbreak\bigbreak
Thus
\begin{equation}
{dP\over dx}(a_j;a)=-\pi^2 q_j - \sum_k(-1)^k R(a_j,a_k) p_k\, .
\label{dP/dx}
\end{equation}
Using (\ref{dQ/dx}) and (\ref{dP/dx}) in the expression for the
diagonal of $R$ in Lemma~\ref{lemma:R_representation}, we find
\begin{equation}
R(a_j,a_j)=\pi^2 q_j^2 +p_j^2 + \sum_k (-1)^k R(a_j,a_k)R(a_k,a_j) (a_j-a_k)
\, .
\label{jmms4}
\end{equation}
\par
Using
\[ {\partial q_j\over\partial a_j} = {dQ(x;a)\over dx}\Bigr\vert_{x=a_j}
+ {\partial Q(x;a)\over \partial a_j}\Bigr\vert_{x=a_j}\, ,
\]
(\ref{jmms2}) and (\ref{dQ/dx}) we obtain
\begin{equation}
{\partial q_j\over \partial a_j} = p_j - \sum_{k\neq j} (-1)^k
R(a_j,a_k) q_k \, .
\label{jmms5}
\end{equation}
Similarly,
\begin{equation}
{\partial p_j\over \partial a_j} = -\pi^2 q_j - \sum_{k\neq k} (-1)^k
R(a_j,a_k) p_k\, .
\label{jmms6}
\end{equation}
\par
Equations (\ref{jmms1})--(\ref{jmms3}) and (\ref{jmms4})--(\ref{jmms6})
are the JMMS equations.  We remark that they appear in slightly
different form in~\cite{jmms} due to the use of sines and cosines
rather than exponentials in the definitions of $Q$ and $P$.
\subsection{Hamiltonian Structure of the JMMS Equations}
\label{subsec:hamiltonian_jmms}
To facilitate comparison with~\cite{jmms,moser} we introduce
\begin{eqnarray}
q_{2j}=-{i\over 2}x_{2j}\, , & \ \ \ &  q_{2j+1}={1\over 2} x_{2j+1}\,
 ,\nonumber\\
p_{2j}=-iy_{2j}\, , &\ \ \ &  p_{2j+1}=y_{2j+1}\, ,\nonumber \\
M_{jk}&:=&{1\over 2}(x_j y_k-x_k y_j)\, , \nonumber \\
G_j(x,y)&:=&{\pi^2\over 4} x_j^2+y_j^2-\sum_{k=1\atop k\neq j}^{2m}
{M_{jk}^2\over a_j-a_k}\, .
\end{eqnarray}
In this notation,
\[ \omega(a)=\sum_j G_j(x,y)\, da_j\, . \]
\par
If we introduce the canonical symplectic structure
\begin{equation}
\left\lbrace x_j,x_k\right\rbrace = \left\lbrace y_j,y_k\right\rbrace=0, \ \ \
\left\lbrace x_j,y_k \right\rbrace = \delta_{jk}\, ,
\label{symplectic_structure}
\end{equation}
then as shown in Moser~\cite{moser}
\begin{theorem}
\label{G_involution}
The integrals $G_j(x,y)$ are in involution; that is, if we define the
symplectic
structure by (\ref{symplectic_structure}) we have
\[ \left\lbrace G_j,G_k\right\rbrace = 0 \ \ \ \mbox{\rm for all}\ \ \
  j,k=1,\ldots,2m.\]
\end{theorem}
Furthermore as can be easily
verified, the JMMS equations take the following form~\cite{jmms}:
\begin{theorem}
\label{jmms_hamiltonian}
If we define the Hamiltonian
\[ \omega(a)= \sum_{j=1}^{2m} G_j(x,y) da_j\, , \]
then Eqs.~(\ref{jmms2}),  (\ref{jmms3}), (\ref{jmms5}) and (\ref{jmms6})
are equivalent to
Hamilton's equations
\[ d_a x_j=\left\lbrace x_j,\omega(a)\right\rbrace \ \ \ \mbox{\rm and} \ \ \
d_a y_j =\left\lbrace y_j,\omega(a)\right\rbrace\, . \]
\end{theorem}
In words, the flow of the point $(x,y)$ in the ``time variable''
$a_j$ is given by Hamilton's equations with Hamiltonian $G_j$.
\par
The (Frobenius) complete integrability of the JMMS equations follows
immediately
from Theorems~\ref{G_involution} and \ref{jmms_hamiltonian}.  We must show
\[ d_a\left\lbrace x_j,\omega\right\rbrace=0 \ \ \ \mbox{\rm and} \ \ \
d_a \left\lbrace y_j,\omega\right\rbrace=0.
\]
Now
\[ d_a\left\lbrace x_j,\omega\right\rbrace=\left\lbrace d_a
x_j,\omega\right\rbrace +
\left\lbrace x_j,d_a\omega\right\rbrace\,,
\]
but $d_a \omega =0 $ since the $G_k$'s are in involution.  And we have
\[\left\lbrace d_a x_j,\omega\right\rbrace=
\sum_{k<\ell}\left(\left\lbrace\left\lbrace x_j,G_k\right\rbrace
,G_\ell\right\rbrace -
\left\lbrace\left\lbrace x_j,G_\ell\right\rbrace ,G_k\right\rbrace\right)
da_k\wedge da_\ell
\]
which is seen to be zero from Jacobi's identity and the involutive property of
the $G_j$'s.
\bigbreak\bigbreak
\subsection{Reduction to Painlev{\'e} V in the One Interval Case }
\label{subsec:p5}
\subsubsection{The $\sigma(x;\lambda)$ differential equation}
\label{subsubsec:sigma}
We consider the  case of one interval:
\begin{equation}
m=1,\ \ \ a_1=-t, \ \ \ a_2=t, \ \ \ \mbox{\rm with}\ \ \ s:=2t\, .
\end{equation}
Since $\rho(x,y)$ is both symmetric and even for $x,y\in I$, we
have $q_2=-q_1$ and $p_2=p_1$.  Introducing the quantity
\[ r_1 = \int_{-t}^t \rho(-t,x)\exp(-i\pi x)\, dx\, , \]
we write
\[ q_1= {\sqrt\lambda\over 2\pi i}(\overline{r_1}-r_1)\ \ \ {\rm and}
\ \ \ p_1={\sqrt{\lambda}\over 2}(\overline{r_1}+r_1)\, . \]
Specializing the results of
Sec.~\ref{subsec:jmms_derivation} to $m=1$ we have
\begin{eqnarray}
\omega(a)&=&-2 R(t,t)\, dt\> , \label{eq1} \\
R(-t,t)&=& -{1\over t} q_1 p_1 =
 {\lambda\over 4\pi i t}(r_1^2-{\overline r_1^2})\, ,
\label{eq2} \\
{dq_1\over dt}&=&-{\partial q_1\over \partial a_1}+{\partial q_1\over \partial
a_2}= - p_1 + 2 R(-t,t) q_1\,  , \nonumber  \\
{dp_1\over dt}&=& \pi^2 q_1+ 2R(-t,t) p_1\, , \nonumber \\
{dr_1\over dt}&=& i\pi r_1 + 2 R(-t,t) {\overline r_1}\, ,  \label{eq3}\\
R(t,t)&=&\pi^2q_1^2+p_1^2-2t R(-t,t)^2 \nonumber \\
&=& \lambda {\overline r_1} r_1 +{\lambda^2\over 8\pi^2 t}\left(
{\overline r_1^2}-r_1^2\right)^2\, . \label{eq4}
\end{eqnarray}
A straightforward computation from (\ref{eq2})--(\ref{eq4}) shows
\begin{eqnarray}
{d\over dt} \left(t R(-t,t)\right) &=& \lambda \Re(r_1^2)\> , \label{eq5}\\
{d\over dt} \left(t R(t,t) \right)&=& \lambda \bigl\vert r_1\bigr\vert^2
\> , \label{eq6}\\
{d\over dt} R(t,t)& =& 2 \left(R(-t,t)\right)^2\> . \label{eq7}
\end{eqnarray}
Eq.~(\ref{eq7}) is known as {\it Gaudin's relation\/}
 and Eqs.~(\ref{eq5}) and
(\ref{eq6}) are identities derived by  Mehta~\cite{mehta92b}
(see also Dyson\cite{dyson92})
 in his    proof of the one interval JMMS equations.
Here we made the JMMS equations  central and derived
(\ref{eq5})--(\ref{eq7})
 as consequences.
\par
These equations  make it easy to derive a
 differential equation for
\begin{equation}
\sigma(x;\lambda):= -2t R(t,t)= x{d\over dx} \log D({x\over\pi};\lambda),
 \ \ \ \mbox{\rm where} \ \ \ x=2\pi t.
\end{equation}
We start with the identity
\[  \bigl\vert r_1\bigr\vert^4 = \left(\Re(r_1^2)\right)^2+
\left(\Im(r_1^2)\right)^2\, ,  \]
and define temporarily
$a(t):=t R(t,t)$ and $b(t):=t R(-t,t)$; then (\ref{eq2}),
(\ref{eq5}), and (\ref{eq6}) imply
\[  \left({da\over dt}\right)^2 = \left({db\over dt}\right)^2
+4\pi^2 b^2 \, . \]
Using (\ref{eq7}) and its derivative to eliminate
$b$ and $db/dt$, we get an equation for $a(t)$
and hence $\sigma(x;\lambda)$:
\begin{theorem}
\label{theorem:p5_representation}
In the case of a single interval $I=(-t,t)$ with $s=2t$, the
Fredholm determinant
\[ D(s;\lambda)=\det\left( 1-\lambda K\right)\]
is given by
\begin{equation}
D(s;\lambda)=\exp\left(\int_0^{\pi s} {\sigma(x;\lambda)\over x}\, dx
\right)\> ,
\label{p5_repr1}
\end{equation}
where $\sigma(x;\lambda)$ satisfies the differential equation
\begin{equation}
\left(x\sigma^{\prime\prime}\right)^2
+4\left(x\sigma^\prime-\sigma\right)\left(x\sigma^\prime-\sigma+
(\sigma^\prime)^2\right)=0
\label{sigma_de}
\end{equation}
with boundary condition as $x\rightarrow 0$
\begin{equation}
\sigma(x;\lambda) = -{\lambda\over \pi} x - ({\lambda\over\pi})^2 x^2
- \cdots \> .
\label{sigma_bc}
\end{equation}
\end{theorem}
Proof: Only (\ref{sigma_bc}) needs explanation.  The small $x$ expansion
of $\sigma(x;\lambda)$ is fixed from the small $s$
expansion  of $D(s;\lambda)$
which can be computed from the Neumann  expansion.\qed
\par
The differential equation (\ref{sigma_de}) is the ``$\sigma$ representation''
of the Painlev{\'e} V equation.  This is discussed in~\cite{jmms}, and in
more detail in  Appendix C of \cite{jimbo_miwa81}.  In terms of the
monodromy parameters $\theta_i$ ($i=0,1,\infty$)
of~\cite{jimbo_miwa81}, (\ref{sigma_de})
corresponds to $\theta_0=\theta_1=\theta_\infty=0$ which is the
case of no local monodromy.  For an introduction to Painlev{\'e} functions
see~\cite{gauss_painleve,levi_winternitz}.
\subsubsection{The $\sigma_\pm(x;\lambda)$ equations}
\label{subsubsec:sigma_pm}
Recalling the discussion following Eq.~(\ref{det_pm}), we see we
need the determinants $D_\pm(s;\lambda)$ to compute
$E_\beta(n;s)$ for  $\beta=1$ and 4.
Let
$R_\pm$ denote the resolvent kernels for the operators
\[ K_\pm := {1\pm J\over 2} K = K {1\pm J\over 2} \> , \]
where $(J f)(x)=f(-x)$ and the last equality  of the above equation
follows from the evenness of $K$. Thus
\[ R_\pm:=(1-K_\pm)^{-1} K_\pm = {1\over 2}(1\pm J) R\> , \] which
in terms of kernels is
\[ R_\pm(x,y)={1\over 2}\left( R(x,y) \pm R(-x,y) \right )\, . \]
Thus
\[ \left(R_-(t,t)-R_+(t,t)\right)^2 = \left(R(-t,t)\right)^2=
{1\over 2} {d\over dt} R(t,t) \, .\]
Introducing the analogue  of $\sigma(x;\lambda)$, i.e.
\[ \sigma_\pm (x;\lambda):= x {d\over dx} \log D_\pm ({x\over\pi};\lambda),\]
the above equation becomes
\begin{equation}
\left({\sigma_-(x;\lambda)-\sigma_+(x;\lambda)\over x}\right)^2
= - {d\over dx}{\sigma(x;\lambda)\over x} \> .
\label{sigma_relations1}
\end{equation}
Of course, $\sigma_\pm(x;\lambda)$  also satisfy
\begin{equation}
 \sigma_+(x;\lambda)+\sigma_-(x;\lambda)=\sigma(x;\lambda)\>  .
\label{sigma_relations2}
\end{equation}
Using (\ref{sigma_relations1}) and (\ref{sigma_relations2})
 and integrating $\sigma_\pm(x;\lambda)/x$ we
obtain (the square root sign ambiguity can be fixed from
small $x$ expansions)
\begin{theorem}
\label{theorem:p5_representation2}
Let $D_\pm(s;\lambda)$ be the Fredholm determinants defined
by (\ref{det_pm}), then
\begin{equation}
\log D_\pm(s;\lambda)={1\over 2}\log D(s;\lambda) \pm {1\over 2}
\int_0^s \sqrt{-{d^2\over dx^2}\log D(x;\lambda)}\, dx\> .
\label{p5_repr2} \end{equation}
\end{theorem}

\section{ASYMPTOTICS}
\label{sec:asym}
\subsection{Asymptotics via the Painlev{\'e} V  Representation}
\label{subsec:asy_p5}
In this section we explain how one derives asymptotic formulas
for $E_\beta(n;s)$ as  $s\rightarrow\infty$ starting with the
Painlev{\'e} V representations of Theorems \ref{theorem:p5_representation}
and \ref{theorem:p5_representation2}.  This section follows
Basor, Tracy and Widom~\cite{btw} (see also \cite{mehta_mahoux}).
We remark that the asymptotics of $E_\beta(0;s)$
as $s\rightarrow\infty$  was first derived
by Dyson~\cite{dyson76} by a clever use of inverse scattering methods.
\par
 Referring to Theorems \ref{theorem:p5_representation} and
\ref{theorem:p5_representation2} one sees that the basic problem
from the differential equation point of view is to derive large $x$
expansions for
\begin{eqnarray}
\sigma_0(x)&:=& \sigma(x;1) \label{sigma0} \\
\sigma_n(x)&:=& {\partial^n\sigma\over \partial \lambda^n}(x;1)\, , \ \ \
n=1,2,\ldots \label{sigma_n} \\
\sigma_{\pm,n}(x)&:=& {\partial^n\sigma_\pm \over \partial \lambda^n}(x;1)\, ,
\ \ \
n=1,2,\ldots \, . \label{sigmapm_n}
\end{eqnarray}
We  point out the
sensitivity of these  results  to the parameter $\lambda$ being set to one.
This dependence is best discussed in terms of the
differential equation
 (\ref{sigma_de}) where it is an instance  of the general problem of
  {\it connection
formulas\/}, see e.g.~\cite{levi_winternitz} and references therein.
 In this context the problem is:  given the
 small $x$ boundary condition,
   find asymptotic formulas as $x\rightarrow\infty$
where all constants not determined by a local analysis
at $\infty$ are given
as functions of the parameter $\lambda$.  If we assume
an asymptotic  solution
for large $x$ of the form $\sigma(x) \sim a  x^p $, then (\ref{sigma_de})
implies  either $p=1$ or $2$ and if $p=2$ then necessarily
$a=-\frac{1}{4}$.  The connection
problem for (\ref{sigma_de})  has been
 studied  by McCoy and Tang~\cite{mccoy_tang}
who show  that    for $0<\lambda < 1$ one has
\[ \sigma(x,\lambda)=a(\lambda) x + b(\lambda) + \mbox{o}(1) \]
as $x\rightarrow\infty$ with
\[ a(\lambda)= {1\over \pi}\log(1-\lambda)
 \ \ {\rm and} \ \    b(\lambda)=
{1\over 2}a^2(\lambda)\, . \]
  Since these formulas make no sense at
$\lambda=1$,
 it is reasonable to guess that
\begin{equation}
 \sigma(x;1)\sim -{1\over 4} x^2\, .
\label{sigma0_asy1}
\end{equation}
For a rigorous proof of this fact see~\cite{widom92}.  It should be noted
that in Dyson's work he too ``guesses'' this leading behavior to
get his asymptotics to work (this leading behavior is not unexpected
from the continuum Coulomb gas approximation).
Given (\ref{sigma0_asy1}),
 and only this, it is
 a simple matter using (\ref{sigma_de}) to compute recursively the  correction
terms  to this  leading asymptotic behavior:
\begin{equation}
\sigma_0(x)=-\frac{1}{4} x^2 - \frac{1}{4} + \sum_{n=1}^\infty {c_{2n}
\over
x^{2n}}\, , \  \ \ x\rightarrow\infty
\label{sigma0_asy2}
\end{equation}
($c_2=-\frac{1}{4}$, $c_4=-\frac{5}{2}$,  etc.).
  Using (\ref{sigma0_asy2})  in
(\ref{p5_repr1}) and (\ref{p5_repr2})
 one can  efficiently generate the  large $s$
expansions  for $D(s;1)$
and $D_\pm(s;1)$  except
for  overall multiplicative constants.  In this instance
other methods fix  these constants (see discussion in~\cite{btw,mehta_book}).
In general this overall multiplicative constant in a $\tau$-function
is quite difficult to determine (for an example of such a determination
see~\cite{bt1,bt2}).
We record here the result:
\begin{equation}
\log D_\pm(s;1)=-\frac{1}{16} \pi^2 s^2\mp \frac{1}{4} \pi s
-\frac{1}{8}\log\pi s
 \pm \frac{1}{4}\log 2 +
\frac{1}{12}\log 2 +
\frac{3}{2}\zeta^\prime(-1) + \mbox{o}(1)\label{logDPM}
\end{equation}
as $s\rightarrow\infty$
where $\zeta$ is the Riemann zeta function.
 We mention
that for  $0<\lambda<1$ the
asymptotics of $D(s;\lambda)$
as $s\rightarrow\infty$   are known~\cite{bw83,mccoy_tang}.

\par
One method to determine the asymptotics of $\sigma_n(x)$
as $x\rightarrow\infty$  is to
examine the variational  equations of (\ref{sigma_de}), i.e.\
simply differentiate (\ref{sigma_de}) with respect to $\lambda$ and
then set $\lambda=1$.  These linear differential equations can
be solved asymptotically given the asymptotic solution (\ref{sigma0_asy2}).
In carrying this out one finds there are two undetermined constants
corresponding to the two linearly independent solutions to the first
variational equation of (\ref{sigma_de}).  One constant does not affect
the asymptotics of $\sigma_1(x)$ (assuming the other is nonzero!).
Determining these constants is part of the general connection problem and
it has not been solved in the context of differential equations.  In
\cite{btw,widom92} Toeplitz and Wiener-Hopf methods are employed to fix
these constants.
The Toeplitz arguments of~\cite{btw} depend upon some unproved
assumptions about scaling limits, but the considerations of
\cite{widom92} are completely rigorous and we deduce  the following
 result~\cite{btw}
for $\sigma_n(x)$  for all $n=3,4,\ldots$:
\begin{equation}
\sigma_n(x)= -{ n!\over (2^3\pi)^{n/2}}{\exp(nx)\over x^{n/2 -1}}
\biggl [1+\frac{1}{8}(7n-4){1\over x}+
\frac{7}{128} (7n^2+12n-16){1\over x^2} +
\mbox{O}({1\over x^3})\biggr ]\label{sigN}
\end{equation}
as $x\rightarrow\infty$.
For $n=1,2$ the above is correct for the leading behavior but
for $n=1$ the correction terms
have coefficients  $\frac{5}{8}$ and $\frac{65}{128}$,
 respectively, and for $n=2$
the above formula gives the  coefficient for $1/x$ but the coefficient
for $1/x^2$ is $\frac{65}{32}$.  See~\cite{btw} for asymptotic formulas
for $\sigma_{\pm,n}(x)$.
\par
Since the asymptotics of $E_\beta(0;s)$ are known, it is convenient
to introduce
\[ r_\beta(n;s):= {E_\beta(n;s)\over E_\beta(0;s)} \, .\]
Here we restrict our discussion to $\beta=2$ (see~\cite{btw} for
other cases).  Using (\ref{sigN}) in (\ref{level_spacing_distr})
one discovers that there is a great deal of cancellation in the
terms which go into the asymptotics of $r_2(n;s)$.  To prove a
result for all  $n\in {\bf N}$ by this method we must
handle all the  correction terms in  (\ref{sigN})---this
was not done in~\cite{btw} and so the following result was proved
only for $1\leq n \leq 10$:
\begin{equation}
r_2(n;s)=B_{2,n}{\exp (n\pi s)\over s^{n^2/2}}\biggl [
1+{n\over 8}(2n^2+7)\, {1\over\pi s} +
 {n^2\over 128}(4n^4+48 n^2+229)\,
{1\over (\pi s)^2}+\mbox{O}({1\over s^3})\biggr ]\label{r2}
\end{equation}
where
\[ B_{2,n}=2^{-n^2-n/2}\pi^{-(n^2+n)/2}\, (n-1)!\, (n-2)!\, \cdots\,
2!\,   1!\; . \]
In the next section we derive the leading term of (\ref{r2})
for all $n\in {\bf N}$.
Asymptotic formulas for $r_\beta(n;s)$
($\beta=1,4,\pm$)   can be
found in~\cite{btw}.

\subsection{Asymptotics of $r_2(n;s)$ from Asymptotics of Eigenvalues}
\label{subsec:asy_eig}
The asymptotic formula (\ref{r2}) can also be derived by a completely
different method (as was briefly indicated in~\cite{btw}).
If we denote
the eigenvalues of the integral operator $K$ by
$\lambda_0>\lambda_1>\cdots >0 $,  then
\[ {\rm det}\left(1-\lambda K\right ) = \prod_{i=0}^\infty (1-\lambda
\lambda_i)\, , \]
and so it follows immediately from (\ref{level_spacing_distr}) that
\begin{equation}
r_2(n;s) = \sum_{i_1<\cdots < i_n} {\lambda_{i_1}\cdots \lambda_{i_n}
\over (1-\lambda_{i_1})\cdots (1-\lambda_{i_n})} \, .
\label{r2_sum}
\end{equation}
(This is formula (5.4.30) in~\cite{mehta_book}.)\ \    Thus the
asymptotics of the eigenvalues $\lambda_i$ as $s\rightarrow \infty$
can be expected to give information on the asymptotics of
$r_2(n;s)$ as $s\rightarrow\infty$.
\par
It is a remarkable fact that the integral operator $K$, acting
on the interval $(-t,t)$, commutes with the differential operator
${\cal L}$ defined by
\begin{equation}
{\cal L}f = {d\over dx} (x^2-t^2) {df\over dx} + t^2 x^2 f,\ \ \ s=2t;
\label{L_operator}
\end{equation}
the boundary condition here is that $f$ be continuous at $\pm t$.
Thus the integral operator and the differential operator have
precisely the same eigenfunctions---the so-called prolate
spheroidal wave functions, see e.g.~\cite{meixner_schafke}.
Now Fuchs \cite{fuchs64}, by an application of the WKB method
to the differential equation, and using a connection between
the eigenvalues $\lambda_i$ and the values of the normalized
eigenfunctions at the end-points, derived the asymptotic
formula
\begin{equation}
1-\lambda_i\sim \pi^{i+1} 2^{2i+3/2} s^{i+1/2} e^{-\pi s}/i!
\label{eigenvalues_asy}
\end{equation}
valid for fixed $i$ as $s\rightarrow\infty$.  Further terms
of the asymptotic expansion for the ratio of the two sides were
obtained by Slepian~\cite{slepian}.
\par
If one looks at the asymptotics of the individual terms on the
right side of (\ref{r2_sum}), then we see from (\ref{eigenvalues_asy})
that they all have the exponential factor $e^{n\pi s}$ and that the
powers of $s$ that occur are
\[ s^{-n/2 - (i_1+\cdots+i_n)}\, . \]
Thus the term corresponding to $i_1=0$, $i_2=1$,\ \ldots, $i_n=n-1$
dominates each of the others.  In fact we claim  this term dominates
the sum of all the others, and so
\begin{equation}
r_2(n;s)\sim 1!\,  2! \cdots (n-1)!\,  \pi^{-n(n+1)/2} 2^{-n^2-n/2}
s^{-n^2/2} e^{n\pi s}\, ,
\end{equation}
in agreement with (\ref{r2}).
\par
To prove this claim, we write
\[ r_2(n;s)= {\lambda_{i_1^0}\cdots \lambda_{i_n^0}\over
(1-\lambda_{i_1^0})\cdots (1-\lambda_{i_n^0})} +
\sum\nolimits^\prime {\lambda_{i_1}\cdots\lambda_{i_n}\over
 (1-\lambda_{i_1})\cdots
(1-\lambda_{i_n})}\]
where $i_1^0=0$, $i_1^0=1$,\ \ldots, $i_n^0=n-1$ and the sum
here is taken over all $(i_1,\ldots,i_n)\neq (i_1^0,\ldots,i_n^0)$
with $i_1<\cdots < i_n$.  We have to show that
\[ \sum\nolimits^\prime
 {\lambda_{i_1}\cdots\lambda_{i_n}\over (1-\lambda_{i_1})\cdots
(1-\lambda_{i_n})}\Bigr/ {e^{n\pi s}\over s^{n^2/2}}\longrightarrow 0 \]
as $s\rightarrow\infty$ and we know that this would be true if the
sum were replaced by any summand.  Write $\sum^\prime = \sum\nolimits_1 +
 \sum\nolimits_2$
where in $\sum\nolimits_1$ we have $i_j<N$ for all $j$ ($N$ to be determined
later) and $\sum\nolimits_2$ is the rest of $\sum\nolimits^\prime$.
Since $\sum\nolimits_1$ is a finite sum we have
\[ \sum\nolimits_1\Bigr/ {e^{n\pi s}\over s^{n^2/2}}\longrightarrow 0 \]
so we need consider only $\sum\nolimits_2$. In any summand
of this we have $i_j\geq N$ for some $j$, and so for this $j$
\[ {1\over 1-\lambda_{i_j}} \leq {1\over 1-\lambda_N} \]
and by (\ref{eigenvalues_asy}) this is at most
\[ a_N\,  s^{-N-1/2}\,  e^{\pi s} \]
for some constant $a_N$.  The product of all other factors
\[  {1\over 1-\lambda_{i_j}} \]
appearing in this summand is at most
\[ \left ( {1\over 1-\lambda_0}\right )^{n-1} \]
and so by (\ref{eigenvalues_asy}) with $i=0$ at most
\[ b_n\,  s^{-(n-1)/2}\,  e^{(n-1)\pi s} \]
for another constant $b_n$.  So we have the estimate
\[ \sum\nolimits_2 \leq a_N b_n\,  s^{-N-n/2}\,  e^{n\pi s}
\sum \lambda_{i_1}\cdots \lambda_{i_n} \]
where the sum on the right may be taken over all $n$-tuples
$(i_1,\ldots,i_n)$.  This sum is precisely equal to
$({\rm tr}\, K )^n=s^n$.  Hence
\[ \sum\nolimits_2\leq a_N b_n s^{-N+n/2} e^{n\pi s}\, .\]
If we choose $N> (n^2+n)/2$ then we have
\[ \sum\nolimits_2\Bigr/ {e^{n\pi s}\over s^{n^2/2}}\longrightarrow 0 \]
as desired.
\vfill\eject
\subsection{Dyson's Continuum Model}
In~\cite{dyson92} Dyson     constructs a
continuum   Coulomb gas  model~\cite{mehta_book}
 for $E_\beta(n;s)$.  In this continuum model,
\[ E_\beta(n;s)=\exp\left(-\beta W- (1-{\beta\over 2})S\right)\]
where
\[ W= -{1\over 2}\int\int \hat\rho(x)\hat\rho(y) \log\vert x-y\vert
\,dxdy\]
is the total energy,
\[  S=\int \rho(x) \, \log\rho(x) \, dx\]
is the entropy,
 $\hat\rho(x)=\rho(x)-1$ and $\rho(x)$ is a continuum
charge distribution on the line satisfying $\rho(x)\rightarrow 1$
as $x\rightarrow\pm\infty$ and $\rho(x)\geq 0$ everywhere.  The distribution
$\rho(x)$ is chosen to minimize the free energy  subject to the condition
\[  \int_{-s/2}^{s/2}\rho(x)\, dx = n\, . \]
 Analyzing his
solution
in the limit $1<<n<<s$,  Dyson finds $E_\beta(n;s)\sim\exp(-\beta W_c)$
where
\begin{equation}
W_c= {\pi^2 s^2\over 16} - {\pi s\over 2}(n+\delta) + {1\over 4}n(n+\delta)
+{1\over 4}n(n+2\delta)\Bigl[\log\bigl({4\pi s\over n}\bigr)+
{1\over 2}\Bigr] \label{dyson}
\end{equation}
with $\delta=1/2-1/\beta$.
\par
We now compare these predictions of the continuum model with the
exact results.  First of all, this continuum prediction does not
get the $s^{-1/4}$ (for $\beta=2$)
or the $s^{-1/8}$ (for $\beta=1,4$)
 present in all $E_\beta(n;s)$ that come from the
$\log\pi s$ term in (\ref{logDPM}).
Thus it is better to compare with  the continuum prediction for $r_\beta(n;s)$.
We find that the continuum model gives both the correct exponential
behavior {\it and\/} the correct power of $s$ for all three ensembles.
Tracing Dyson's arguments shows that
 the power of $s$ involving  the
$n^2$ exponent is an {\it energy effect\/} and the power of $s$
involving  the
$n$ exponent is  an {\it entropy effect\/}.
Finally, the continuum model also makes a prediction (for large $n$) for
the the $B_{\beta,n}$'s ($B_{2,n}$ is given
above).    Here we find that the ratio
of the exact result to the continuum model result is approximately
$n^{-1/12}$ for $\beta=2$ and $n^{-1/24}$ for $\beta=1,4$.  This
prediction of the continuum model is better than it first appears when
one considers that the constants themselves are of order
$n^{n^2/2}$ ($\beta=2$) and $n^{n^2}$ ($\beta=1,4$).
\vfill\eject
\acknowledgments
It is a pleasure to acknowledge E.~L.~Basor, F.~J.~Dyson, P.~J.~Forrester,
 M.~L.~Mehta, and P.~Nevai  for their many helpful comments and
their  encouragement.  We also thank F.~J.~Dyson and M.~L.~Mehta
for sending us their preprints prior to publication.  The first
author  thanks  the organizers of the $8^{th}$ Scheveningen
Conference, August 16--21, 1992,  for the invitation to attend
and  speak at this conference.  These notes are an expanded version
of the lectures presented there.  This work was supported in part
by the National Science Foundation, DMS--9001794 and DMS--9216203,
and this support is gratefully acknowledged.

\vfill\eject
\figure{Density of eigenvalues histogram for 25, $100\times 100$ GOE
matrices.  Also plotted is the Wigner semicircle which is known to be
the limiting distribution.}
\figure{Density of eigenvalues histogram for 25, $100\times 100$
symmetric matrices whose elements are uniformly
distributed
on $[-1,1]$.  Also plotted is the Wigner semicircle distribution.}
\figure{Level spacing histogram for 20, $100\times 100$ GOE
matrices.  Also plotted is the Wigner surmise (\ref{wigner}).}
\figure{Level spacing histogram for 50, $100\times 100$ symmetric
matrices whose elements are uniformly distributed on $[-1,1]$.
Also plotted is the Wigner surmise (\ref{wigner}).}
\figure{Level spacing histogram for mixed data sets.  Also plotted
is the Poisson distribution.}
\end{document}